\documentclass[letterpaper]{article} %
\usepackage{aaai23}  %
\usepackage{times}  %
\usepackage{helvet}  %
\usepackage{courier}  %
\usepackage[hyphens]{url}  %
\usepackage{graphicx} %
\usepackage{natbib}  %
\usepackage{caption} %
\usepackage{algorithm}
\usepackage{algorithmic}

\usepackage{newfloat}
\usepackage{listings}
\DeclareCaptionStyle{ruled}{labelfont=normalfont,labelsep=colon,strut=off} %
\floatstyle{ruled}
\newfloat{listing}{tb}{lst}{}
\floatname{listing}{Listing}
\usepackage{amsmath,amsfonts,bm}

\def\Figref#1{Figure~\ref{#1}}

\def\Tabref#1{Table~\ref{#1}}

\def\Secref#1{Section~\ref{#1}}

\def\Appref#1{Appendix~\ref{#1}}

\def\eqref#1{equation~\ref{#1}}

\def\1{\bm{1}}

\DeclareMathAlphabet{\mathsfit}{\encodingdefault}{\sfdefault}{m}{sl}
\SetMathAlphabet{\mathsfit}{bold}{\encodingdefault}{\sfdefault}{bx}{n}

\newif\ifaaai
\newif\ifamlc
\aaaitrue

\usepackage{comment}
\usepackage{multirow}
\usepackage{booktabs}
\usepackage{graphicx}
\usepackage{adjustbox}
\usepackage{url}

\usepackage{ulem} %
\usepackage[inline]{enumitem} %

\usepackage{xcolor}
\usepackage{listings}

\usepackage{xparse}

\newcommand{\codeinline}[1]{\lstinline{#1}}
\NewDocumentCommand{\codeword}{v}{%
\texttt{\textcolor{blue}{#1}}%
}

\renewcommand{\emph}[1]{\textit{#1}}

\def\ContinueLineNumber{\lstset{firstnumber=last}}

\usepackage{xspace} %
\def\CallArgs{\textsc{CallArgs}\xspace}
\def\PyEnvs{\textsc{PyEnvs}\xspace}

\title{Better Context Makes Better Code Language Models: \\
A Case Study on Function Call Argument Completion}
\author{
    Hengzhi Pei\textsuperscript{\rm 1}\thanks{Work done while interning at Amazon Web Services.},
    Jinman Zhao\textsuperscript{\rm 2},
    Leonard Lausen\textsuperscript{\rm 2},
    Sheng Zha\textsuperscript{\rm 2},
    George Karypis\textsuperscript{\rm 2}
}
\affiliations{
    \textsuperscript{\rm 1} University of Illinois Urbana-Champaign\\ 
    \textsuperscript{\rm 2} Amazon Web Services\\

    hpei4@illinois.edu, 
    \{jinmaz,lausen,zhasheng,gkarypis\}@amazon.com
}

\begin{document}

\maketitle

\begin{abstract}
Pretrained code language models have enabled great progress towards program synthesis. However, common approaches only consider in-file local context and thus miss information and constraints imposed by other parts of the codebase and its external dependencies. Existing code completion benchmarks also lack such context. To resolve these restrictions we curate a new dataset of permissively licensed Python packages that includes full projects and their dependencies and provide tools to extract non-local information with the help of program analyzers. We then focus on the task of function call argument completion which requires predicting the arguments to function calls. We show that existing code completion models do not yield good results on our completion task. To better solve this task, we query a program analyzer for information relevant to a given function call, and consider ways to provide the analyzer results to different code completion models during inference and training. Our experiments show that providing access to the function implementation and function usages greatly improves the argument completion performance. Our ablation study provides further insights on how different types of information available from the program analyzer and different ways of incorporating the information affect the model performance.

\end{abstract}

\section{Introduction}
\label{sec:introduction}

Following their counterparts in the natural language domain, we see rapid adoption of language models in the code domain, e.g. Code-GPT~\citep{lu2021codexglue}, GPT-J~\citep{wang2021gpt-j}, Codex~\citep{chen2021evaluating}, PLBART~\citep{ahmad2021unified}, CodeT5~\citep{wang2021codet5}, CodeGen~\citep{nijkamp2022conversational}, to name a few.
Such \emph{code language models} have demonstrated amazing abilities in achieving new state of the art in various code-related tasks such as clone detection \citep{feng2020codebert,guo2020graphcodebert, ahmad2021unified, wang2021codet5}, code completion \citep{svyatkovskiy2020intellicode, kim2021code, lu2021codexglue, guo2022unixcoder} and code translation \citep{ahmad2021unified, wang2021codet5, guo2022unixcoder}.
Notably, pretrained large code language models have been reported to solve programming problems at a high rate \cite{chen2021evaluating} or at a rate similar to average human contestants \citep{li2022competition}.

Despite their success, these code language models are usually restricted to file-level local code context, and thus miss richer context from project files and libraries. 
This is in contrast to the more common software development setting where the current piece of code only makes sense under the context and constraints posed by the codebase that it resides in.
As a result, models can generate outputs that appear locally plausible but contradict the constraints imposed by the codebase that they operate in. 

Machine-learning-for-code datasets and benchmarks also lack such project-level environments for training and evaluating code language models.
For example, a common choice for evaluating code completion models is Py150 \citep{raychev2016probabilistic} which contains Python source files from Github. 
However, its train-test split is at file level, making it infeasible to obtain dependency information or incorporate project-level analysis. 
Similar problems applies to other existing datasets in the code domain (more in Section \ref{sec:related-work}). 
This motivates us to create a new dataset that better captures what happens in the real-world where software is developed by incorporating external dependencies and spans multiple files. 
Another missed opportunity is the availability program analyzer smarts that are common in today's integrated development environments (IDEs).
Our dataset provides an extensible foundation for the integration of program analyzer information into machine learning for code models.%

We collect thousands of permissively licensed Python packages and create a development environment for each of them along with their dependencies. 
We utilize a language server which can provide language-specific smarts based on project-level analysis. 
We query the language server to extract environment information relevant to a certain code location. 
Our setup supports extracting all analyzer information that is available to a developer through an IDE, potentially beneficial to a wide range of code-related tasks. 
\ifaaai
\fi

As a case study, we focus on the task of \emph{function call argument completion}, a special case of code completion where we predict what arguments should be passed through a given function call.
This task is well suited to analyze the impact of missing cross-file and cross-project information as understanding function calls requires information that spans file boundaries.
A developer who needs to decide what to pass to a function first needs to understand the meaning and the expected usage.
They may also benefit from looking at other usage examples of the same function, especially those from within the current working project. 
In this regard, we build a task-specific dataset by querying the language server to extract function definition, implementation and usage information for each function call instance.
We show in \Secref{sec:code-completion-experiments} that function call argument completion is more challenging than the average code completion cases and it cannot be made trivial through copying similar occurrences (\Tabref{tab:fc_copy}).

We conduct extensive experiments on feeding function-call-related information to various pretrained code language models for function call argument completion under different settings, with or without further training the models with this additional type of information. 
We concatenate different types of analyzer-induced context together with the original local code context as model inputs.
We evaluated the performance of several state-of-the-art pretrained code language models, including decoder-only models CodeGPT, CodeGPT-adapt, CodeGen, and encoder-decoder models CodeT5, PLBART, UnixCoder. 

We find that providing code language models with analyzer-induced context universally improves their accuracy of call argument prediction.
Both the improvements and the best performances are greater with task-specific fine-tuning.
We also found that under similar settings, encoder-decoder models %
perform better than decoder-only models for this task, which is not the case for general code completion \citep{guo2022unixcoder}. 
See more in \Secref{sec:task-specific-experiments}.
The ablation study (\Secref{sec:ablation-study}) further reveals the unique effects of function implementation and function usage information on model completion performances.

Our main contributions are:
\begin{enumerate*}[label=\itshape\roman*\upshape)]
    \item We collect \PyEnvs, a large number of permissively licensed Python packages along with their isolated development environments which supports querying for project-related information. 
    \item We build \CallArgs, a new dataset for function call argument completion which provides function definition, implementation, and usage information for each function call instance.
    \item We conduct extensive experiments on feeding analyzer-induced code context as input into various pretrained code language models for function call argument completion, and report various findings.
\end{enumerate*}
Our code and data will be made available at \url{https://github.com/amazon-research/function-call-argument-completion}.

\section{Dataset Creation}
\label{sec:dataset}

We describe the formation of \PyEnvs, a sizable collection of permissively licensed Python projects along with their development environments that support queries to an industry-level program analyzer. 
We showcase the creation of \CallArgs, a dataset for function call argument completion, which we use in the later experiments.
The setup can be used to create analyzer-annotated datasets for other code-related tasks.

\subsection{\PyEnvs for Analyzer-Annotated Code Tasks}
\label{sec:dataset-pyenvs}
\ifaaai
We look up the top 5000 most downloaded projects\footnote{https://hugovk.github.io/top-pypi-packages/} from Python Package Index (PyPI). 
To ensure that our environments can be redistributed, we only keep the projects where the project itself and all its dependent packages are permissively licensed\footnote{MIT, Apache, BSD, CC0, ZPL 2.1, ISCL, PSF (Python Software Foundation License), HPND, or Unlicense.}.
For each project we then create a virtual environment\footnote{\url{https://docs.python.org/3/tutorial/venv.html}} including the project code and all its dependencies obtained via the Python Package Installer (\verb|pip|)\footnote{\url{https://github.com/pypa/pip}}. 
For each virtual environment, we locate the source code of the project and its dependencies using the metadata provided by \verb|pip| in the installation directories. 
In all, we obtain 2814 packages along with their virtual environments.
\\
\fi
\ifamlc
We collect 2814 packages out of the top 5000 most downloaded projects from Python Package Index\footnote{https://hugovk.github.io/top-pypi-packages/} (PyPI) where both themselves and their dependencies are permissively licensed\footnote{MIT, Apache, BSD, CC0, ZPL 2.1, ISCL, PSF (Python Software Foundation License), HPND.}.
For each one of them, we create a virtual environment\footnote{\url{https://docs.python.org/3/tutorial/venv.html}} and install the package along with all its dependencies using the Python Package Installer\footnote{\url{https://github.com/pypa/pip}}.
\fi
This setup is reminiscent of a human developer's work environment and enables us and other researchers to build on tools designed for human developers to obtain additional information for machine learning for code models. 
\ifaaai
The Language Server Protocol\footnote{\url{https://microsoft.github.io/language-server-protocol/}} has been proposed to standardize how such tools and IDEs communicate and is commonly used by IDEs such as VSCode.
\fi
For this purpose, we choose the Jedi Language Server\footnote{https://github.com/pappasam/jedi-language-server} based on the popular Jedi autocompletion, static analysis and refactoring library for Python as the program analyzer to extract auxiliary information of each project. It enables us to obtain the location of a function's implementation as well as the locations of other points in the project that use the function.
We set up the language server with project-specific workspaces based on the virtual environments, enabling the language server to provide the same level of information that is accessible by a developer using an IDE.

\subsection{\CallArgs: Function Call Argument Completion Dataset}
\label{sec:dataset-callargs}
\ifaaai
Based on the above, we further construct a new function call argument completion dataset \CallArgs. 
\fi
We parse each Python file in a project into abstract syntactic trees and extract local information related to each function call, which includes: 1) the function name, 2) the location of function arguments, 3) the location of the function call, and 4) the local context's location of the function call.
Here the local context means the function body in which the given function call occurs. In our dataset, we only consider function calls that occur in a function body. 

For analyzer-induced information, we query the language server and get the response about the function definition and signature information of the given function call (details in Appendix \ref{sec:query}). 

We ignore some function calls which we regard as less meaningful for argument predictions. 
For example, function calls related to error messages and logging, function calls which the language server fails to find their definitions, or function calls without any arguments. 
More details about our criteria can be found in Appendix \ref{sec:criteria}.

To prevent information leakage from project dependencies,
we make an effort to ensure both project-level and dependency-level isolation between the training and the test set. 
We propose an isolation strategy for the tasks that access information from direct dependencies of each project such as our function call argument completion task.
We treat the Python standard library built-in packages such as \codeinline{os} and \codeinline{pickle} as public information whose use does not necessitate isolation between training and test set. 
For third-party dependencies (i.e. other projects), we ensure that if a project is a dependency of a project in the training set or itself is in the training set, then it can not be part of the test set or a dependent of any project in the test set; and vice versa.
More details can be found in Appendix \ref{sec:isolation}.

We randomly sample validation and test set
\ifamlc%
\footnote{
Since the same dependency-isolation is not enforced between the validation and the test set, it is possible that information leaks through the validation set. 
However, the strong isolation criterion between the training set and the rest already forced us to exclude 34\% of the 2814 available packages. 
Enforcing the same isolation between the validation and the test will likely result in an even smaller total number of packages.
}%
\fi%
and select training sets to respect the said isolation.
We carried out the process with different sample sizes and chose the resulting split where the ratio among training:validation:test is roughly 10:1:1. 
The statistics is shown in Table \ref{tab:generalstat}. 

\ifaaai
\begin{table}[ht]
\centering
\begin{tabular}{ccccc}
\toprule
{} & No. of  & No. of  & No. of  & No. of  \\
Split & projects & files & tokens & function calls \\
\midrule
Train & 1578 & 13790    & 36.2M  & 364752 \\
Validation   & 145  & 2496     & 5.2M   & 42841  \\
Test  & 145  & 1701     & 2.5M   & 49085  \\   
\bottomrule
\end{tabular}
\caption{Statistics of the \CallArgs dataset. }
\label{tab:generalstat}
\end{table}
\fi

\ifaaai
\fi

\section{Task \& Method}
\label{sec:method}
In this section, we formulate the call-argument-completion task, and describe how we incorporate the static analyzer information as function implementation context and function usage context to code language models. 

\subsection{Task Formulation}

\begin{table}[ht]
\centering
\begin{tabular}{p{0.97\columnwidth}}
\toprule
Example: \codeinline{<PREDICT>} denotes the prediction location.  \\
\midrule
\begin{tabular}[c]{l}
\begin{lstlisting}[]
def _set_arguments(self,arguments,context):
  positional, named = arguments
  variables = context.variables
  args, kwargs = self.arguments.map(
\end{lstlisting}
\end{tabular} \\
\ContinueLineNumber
\begin{tabular}[c]{l}
\begin{lstlisting}
        <PREDICT>
\end{lstlisting}
\end{tabular} \\
\ContinueLineNumber
\begin{tabular}[c]{l}
\begin{lstlisting}[]
  self._set_variables(args, kwargs, variables)
  context.output.trace(lambda: self._trace_log_args_message(variables))
\end{lstlisting}
\end{tabular} \\
\bottomrule
\end{tabular}
\caption{An argument completion example for \codeinline{self.arguments.map()}. For unidirectional prediction, the local context consists of the left context (Line 1-4) only; for in-filling prediction, the local context is given as the left local context (Line 1-4) and right local context (Line 6-7). }
\label{tab:examples}
\end{table}

We define our call argument completion task to be: given the available context (local, project-level, or beyond) of a given function call, predict the complete list of (positional and keyword) arguments to be passed to the function call. 

We treat code as sequences of code tokens.
We start with the base case where only the in-file local context around the function call location is available.
Assume a code-token sequence $X = [x_1, x_2, ..., x_n] $ where $ X_{f} = [x_j, x_{j+1}, ... x_{l-1} ] $ is a function-call occurrence and $X_{a} = [x_l, x_{l+1}, ... x_{r} ] $ is the list of arguments passed to the function call. 
In our task, we restrict $X$ to be a Python function that contains at least one function call. 
We refer to the tokens before the target arguments $ X_L = [x_{i}]_{i<l} $ as the left local context, and the tokens after the target arguments $ X_R = [x_i]_{i>r} $ as the right local context of the function call $X_f$.

We consider two variations of the task.
In \emph{unidirectional prediction}, we model $ P(X_a \mid X_L) $. This is a widely adopted scenario for code completion \citep[e.g. ][]{karampatsis2020big, lu2021codexglue, kim2021code, lu2022reacc}, which simulates the case that a developer is writing code from the beginning to end, where only the local context up to the prediction point is available. 
In \emph{in-filling prediction}, we model $ P(X_a \mid X_L, X_R) $ which simulates the case that a developer is editing an existing piece of source code, where both the left and right context is present.
This setting is similar to cloze tests \cite{feng2020codebert} which aim to predict the token for the blank with the context of the blank. 
Our particular setting is more difficult, as the model needs to continuously generate more than one tokens.
Table \ref{tab:examples} shows an example for the unidirectional prediction and the in-filling prediction.

For both cases, we use a language model to generate predictions for the call arguments.
For unidirectional prediction, a decoder-only model $ P(X_a \mid X_L) = \prod_{i=l}^{r} p(x_i \mid x_{<i}) $ can be used.
For both unidirectional and in-filling prediction, an encoder-decoder model $ P(X_a \mid X_L, X_R) = \prod_{i=l}^{r} p(x_i \mid x_{<i}, X_L, X_R) $ where $ X_L $ and $ X_R $ (if available) are passed into the encoder.

\subsection{Static Analyzer Information}
With the presence of additional information from the static analyzer, we incorporate it as additional context $A$ into the models: $ P(X_a| X_L, A)$ for unidirectional prediction and $ P(X_a| X_L, X_R, A)$ for in-filling prediction.
We use as context the function implementation information and function usage information for function call argument completion.

Function implementation context $Imp$ is the Python function definition of the given function call $X_f$. 
It reveals the formats, the constraints, and the intention of the current function call.  
We retrieve this information with the function definition location from the language server.

Function usage context $\{U_i\}$ collects local contexts surrounding the calls of the same function within the project. 
Specifically, we include the in-file usages that occur before the prediction location and the in-project usages that occur in a different file.
The usage context provides project-specific examples that helps the model to better induce the usage for the current function call. 
We retrieve this context by grouping the function call instances that share the same definition.

Some function calls, especially those defined in the Python standard libraries, can appear many times within the codebase of a project. 
Therefore, we design a similarity-based ranking criteria and only select top usages to provide to a model. 
For a target function call and a usage $u$ of the same call, we calculate the similarity as $ |S_l \cap S_{u}| / |S_l| $ where $S_l$ is the set of local call left context tokens and $S_{u}$ is the set of the left context tokens for $u$.

\subsection{Incorporating Methods}
\label{sec:method-incorporating}

We concatenate analyzer-induced code context with the local context as the model input, which leverages the power of the pretrained language models to implicitly understand and extract information from each part. 
\ifaaai
\Tabref{tab:form} describes the input templates for the decoder-only models and the encoder-decoder models.
\fi
\ifamlc
Specifically, the input template for the decoder-only models is
``\textless{}s\textgreater~{$[Imp]$} \textless{}/s\textgreater~$[U_m]$ \textless{}/s\textgreater~ ... \textless{}/s\textgreater~$[U_1]$ \textless{}/s\textgreater~$[X_L]$''
and the input template for the encoder-decoder models is
``\textless{}s\textgreater~$[X_L]$  \textless{}PREDICT\textgreater~$[X_R]$ \textless{}/s\textgreater~{$[Imp]$} \textless{}/s\textgreater~$[U_1]$ \textless{}/s\textgreater~... \textless{}/s\textgreater~$[U_m]$ \textless{}/s\textgreater{}'',
where 
\textless{}s\textgreater{}, \textless{}/s\textgreater{}, and \textless{}PREDICT\textgreater{} are special tokens, and \textless{}PREDICT\textgreater{} suggests the location to fill in the call arguments.
\fi
We set a length budget for each piece of information so that the total input does not exceed model capacity.
If a context exceeds the allocated length, we drop the function implementation and the right local context from the right, and drop the usage context and the left local context from the left
to preserve the most relevant information.

We consider the case both with or without task-specific fine-tuning. 
We call the former setting \emph{concatenating during inference} (CDI). 
In this setting, the model is never exposed to such auxiliary information.
Since auxiliary information in our case can be efficiently retrieved by the static analyzer, we further consider fine-tuning a model with augmented inputs. 
We hypothesize that this helps a model better understand the relationship between each piece of information.%

\ifaaai

\begin{table*}[ht]
\centering

\begin{tabular}{ll}
\toprule
Type            & Input format                                                                                                                                                                                                                                                              \\
\midrule
Decoder-only    & \textless{}s\textgreater~{$[Imp]$} \textless{}/s\textgreater~$[U_m]$ \textless{}/s\textgreater~ ... \textless{}/s\textgreater~$[U_1]$ \textless{}/s\textgreater~$[X_L]$                                                                        \\
Encoder-Decoder & \textless{}s\textgreater~$[X_L]$  \textless{}PREDICT\textgreater~$[X_R]$ \textless{}/s\textgreater~{$[Imp]$} \textless{}/s\textgreater~$[U_1]$ \textless{}/s\textgreater~... \textless{}/s\textgreater~$[U_m]$ \textless{}/s\textgreater{} \\
\bottomrule
\end{tabular}
\caption{The input formats for decoder-only and encoder-decoder models. \textless{}s\textgreater{}, \textless{}/s\textgreater{}, and \textless{}PREDICT\textgreater{} are special tokens. \textless{}PREDICT\textgreater{} suggests the location to fill in with the call arguments.}
\label{tab:form}
\end{table*}

\fi

\section{Experiment}
In this section, we design experiments to answer the following research questions (RQs).
\begin{enumerate}[label=\textbf{RQ\arabic*},leftmargin=0pt,itemindent=2.5em]
  \item \emph{How do general code completion models perform on function call argument completion?} \\
  We conjecture that general code completion models without project-specific context do not work well on the tasks where external context is critical. 
  To this regard, we test several pretrained code language models on our \CallArgs dataset.
  
  \item \emph{To what extent are the analyzer-induced information helpful for pretrained language models to do function call argument completion?} \\
  To explore this point, we conduct experiments under two settings. We first test if a general code completion model would perform better on our \CallArgs dataset under the CDI setting. Then, we test for unidirectional and in-filling prediction on our \CallArgs dataset after task-specific fine-tuning.
  
  \item \emph{What are the roles and contributions of different types of analyzer-induced information?} \\
  Specifically, we study the impact of function implementation context and function usage context on our \CallArgs dataset. We conduct an ablation study to break down the performance gain from using function implementation context and function usage context. We further explore how the choice of the number and the length of usage context affect model performance. 
\end{enumerate}

\subsection{RQ1}
\label{sec:code-completion-experiments}
To answer RQ1, we evaluate various decoder-only models pretrained on Python source code: CodeGPT~\citep{lu2021codexglue}, CodeGPT-adapted \citep{lu2021codexglue} and CodeGen \citep{nijkamp2022conversational}. 
Since the models are pretrained using different tokenization strategies and different pretraining datasets, 
we fine-tune them using general code completion, aka next-token prediction at \emph{all} tokens, over \CallArgs to reduce the impact from data and domain shift.
Specifically, we follow the preprocessing step in Py150 \citep{raychev2016probabilistic} to standardize the inputs, split the files from \CallArgs training set into code fragments of equal lengths of 1024 tokens, and use the standard causal language modeling loss. 
Project-level information is not presented during this process.

We train for 10 epochs with a batch size of 32 and a learning rate of 5e-5 using AdamW optimizer \citep{loshchilov2018decoupled}. 
We apply early stopping when validation set perplexity does not improve for 3 epochs, and chose the checkpoint with the best validation set perplexity.
We evaluate the token-level accuracy for general code completion on the files from \CallArgs test set. For each argument completion instance, we ask the model to generate arguments from the left local context only (unidirectional prediction).
We use beam search of size 5 until a matching parenthesis is generated. 
We measure the exact match accuracy (EM) and Levensthein edit similarity (EditSim) between the ground truths and the model completions.

The results are shown in Table \ref{tab:autoregressive}. 
We find that, with only local context, although general code completion models can achieve good token-level accuracy (Token-level Acc 72 -- 79), they do not perform well on call argument completion (EM 36 -- 45). 
This suggests that call arguments are more difficult to predict than general locations~\citep[cf.][]{rahman2019natural}.
One possible reason is that general code completion on average involves easier prediction locations such as boiler-plate code and code idioms.

\ifaaai
\begin{table}[ht]
\centering
\resizebox{\columnwidth}{!}{
\begin{tabular}{cccc}
\toprule
 Model (Token-level Acc)   & Context         &  EM & EditSim \\
\midrule
 CodeGPT (72.18) & local context    & 36.29 & 63.50    \\
         & w/ implementation        & 37.40 & 64.75     \\
         & w/ usages                & 44.99  & 73.15   \\
 CodeGPT-adapted (72.59) & local context     & 37.24 & 64.76    \\
                 & w/ implementation         & 38.25 & 65.98     \\
                 & w/ usages                 & 46.58 & 74.36   \\
 UnixCoder-base (75.88) & local context    & 38.45 & 66.11    \\
                 & w/ implementation       & 40.04 & 67.85     \\
                 & w/ usages               & 47.93 & 75.40   \\
 CodeGen-MONO (78.73)  & local context    & 43.45 & 69.69    \\
                 & w/ implementation      & 46.26 & 72.46     \\
                 & w/ usages              & 52.99 & 78.52   \\
\bottomrule
\end{tabular}
}
\caption{The call-argument completion performance of several general code completion models on \CallArgs. 
The token-level accuracy is for general code completion. 
}
\label{tab:autoregressive}
\end{table}
\fi
\ifamlc
\begin{table}[ht]
\centering

\hspace{-1em}
\begin{minipage}{0.4\columnwidth}
\caption{Statistics of the \CallArgs dataset. }
\label{tab:generalstat}

\resizebox{\columnwidth}{!}{
\begin{tabular}{ccccc}
\toprule
{} & No. of  & No. of  & No. of  & No. of  \\
Split & projects & files & tokens & function calls \\
\midrule
Train & 1578 & 13790    & 36.2M  & 364752 \\
Validation   & 145  & 2496     & 5.2M   & 42841  \\
Test  & 145  & 1701     & 2.5M   & 49085  \\   
\bottomrule
\end{tabular}
}
\end{minipage}
\begin{minipage}{0.55\columnwidth}
\caption{The call-argument completion performance of several general code completion models on \CallArgs. 
The token-level accuracy is for general code completion. 
}
\label{tab:autoregressive}
\resizebox{\columnwidth}{!}{
\begin{tabular}{cccc}
\toprule
 Model (Token-level Acc)   & Context         &  EM & EditSim \\
\midrule
 CodeGPT (72.18) & local context    & 36.29 & 63.50    \\
         & w/ implementation        & 37.40 & 64.75     \\
         & w/ usages                & 44.99  & 73.15   \\
 CodeGPT-adapted (72.59) & local context     & 37.24 & 64.76    \\
                 & w/ implementation         & 38.25 & 65.98     \\
                 & w/ usages                 & 46.58 & 74.36   \\
 UnixCoder-base (75.88) & local context    & 38.45 & 66.11    \\
                 & w/ implementation       & 40.04 & 67.85     \\
                 & w/ usages               & 47.93 & 75.40   \\
 CodeGen-MONO (78.73)  & local context    & 43.45 & 69.69    \\
                 & w/ implementation      & 46.26 & 72.46     \\
                 & w/ usages              & 52.99 & 78.52   \\
\bottomrule
\end{tabular}
}
\end{minipage}
\end{table}
\fi

\subsection{RQ2}
\label{sec:task-specific-experiments}
We evaluate the performance of function call argument completion with the presence of analyzer-induced contexts.
When incorporating those contexts, we fill in the respective input slots as described in \Secref{sec:method-incorporating}. 

\subsubsection{Concatenating during inference.}

We directly concatenate analyzer-induced context as input to the same models described in \Secref{sec:code-completion-experiments}. 
We allocate at least a quarter of the total length budget for analyzer-induced context. 
We use no more than 3 function usages for each instance unless otherwise specified.%
The results are shown in Table \ref{tab:autoregressive}. 

We find that both function implementation (w/ implementation) and function usages (w/ usages) universally improve the EM and EditSim across all the models tested, 
indicating that pretrained code language models benefit from auxiliary contexts even without exposure to them during training.
The gains from function usages is much greater (average EM improvements 9.27 vs 1.63), suggesting that if presented only at inference time, similar contexts help code completion more \citep[cf. similar observations from][]{lu2022reacc}.

\subsubsection{Task-specific fine-tuning.}
\label{sec:call-args-completion-results}

Next, we fine-tune different pretrained code language models specifically for call argument completion, with or without the presence of auxiliary contexts.
For decoder-only models, we use CodeGPT \citep{lu2021codexglue}, UnixCoder \citep{guo2022unixcoder} and CodeGen \citep{nijkamp2022conversational}. 
For encoder-decoder models, we use CodeT5 \citep{wang2021codet5}, PLBART \citep{ahmad2021unified} and UnixCoder~\footnote{UnixCoder is pretrained with both denoising objectives and unidirectional language modeling so it can be used as both an encoder-decoder model and a decoder-only model.}~\citep{guo2022unixcoder}. 

We conduct experiments for both the unidirectional and in-filling prediction. 
For the latter, the length budget of the right local context is one-third of the length budget of the left local context. The length budget for function implementation and the average length of each function usage are set as one-eighth of the total input length. 
We train our models for 10 epochs with a batch size of 64 and a learning rate of 2e-5.
We use early stopping based on the perplexity over the validation set.
The results are shown in \Tabref{tab:fc_all}. 

Comparing the unidirectional results from decoder-only models (\Tabref{tab:fc_all}, top section) to those in \Tabref{tab:autoregressive}, we find that task-specific fine-tuning EM and EditSim are on average 4.49 and 5.52 higher across the models using local context only. 
The presence of function implementations and function usages further greatly improves the argument completion performance compared to using local context only, with an average 13.28 gain for EM and an average 8.76 gain for EditSim across all models and settings. 
The best results are achieved by CodeT5-base with 62.73 EM and 84.33 EditSim for unidirectional prediction, and by CodeT5-base with 69.28 EM and 88.08 EditSim for in-filling prediction.

Compared to unidirectional prediction, the in-filling prediction metrics are on average 6.83 and 4.57 higher for EM and EditSim  using the same model and the same source of information. 
We conjecture that this is because the right context reveal the use of the function call result, or because it provides more information for the model to relate to previously seen similar patterns.

For models with similar numbers of trainable parameters, the results from the encoder-decoder models are usually better than the decoder-only models when using the same contexts. 
For example, for unidirectional prediction, the encoder-decoder version of UnixCoder is better than its decoder-only version in both metrics. 
This suggests that the encoder-decoder architecture can be a powerful design for code auto-completion, probably thanks to its better ability at leveraging the input contexts.
The results also suggest that the pretraining objectives are important for model's performance. For example, CodeT5-small (60M) pretrained with mask span prediction is better than PLBART-base (139M) pretrained without it, despite that the former has fewer parameters.

\begin{table*}[ht]
\centering
\resizebox{0.9\linewidth}{!}{
\begin{tabular}{cccccc}
\toprule
Task (model type)      & Model (\# of trainable parameters)    & Context                     & EM   & EditSim & Input length \\
\midrule
Unidirectional & CodeGPT (124M) & local context            &  41.74   & 70.01    & 512     \\
(decoder-only)                &         & +implementation\&usages &  54.19 &  78.88    & 924     \\
  & CodeGPT-adapted (124M) & local context            &  41.65   & 70.05    & 512     \\
                &         & +implementation\&usages &  54.20 &  79.03    & 924     \\
                & UnixCoder-base (126M) & local context               &  42.52 &  71.33   & 512    \\
                &                & +implementation\&usages &  58.22 &  81.46   & 924     \\
                & CodeGen-MONO (355M) & local context          &  47.49  &  74.74   & 512    \\
                &                & +implementation\&usages     & 62.55   &  83.71   & 924     \\
\cmidrule{2-6}
(encoder-decoder)            & CodeT5-small (60M) & local context               & 43.65 & 71.89     & 512                 \\
           &              & +implementation\&usages & 59.16 & 82.34     & 1024                 \\
           & CodeT5-base (223M) & local context               & 47.16 & 74.44     & 512                 \\
           &              & +implementation\&usages & 62.73 & 84.33     & 1024                 \\
           & PLBART-base (139M) & local context               & 38.96 & 68.17     & 512                 \\
           &              & +implementation\&usages & 51.26 & 76.77     & 1024                 \\   
           & UnixCoder-base (126M) & local context            & 46.33 & 73.29     & 512                 \\
           &              & +implementation\&usages & 60.53 & 82.85     & 924                 \\
\midrule
In-filling & CodeT5-small (60M) & local context               & 51.63 & 77.60      & 512                 \\
(decoder-only)            &              & +implementation\&usages & 62.47 & 85.07     & 1024                 \\
           & CodeT5-base (223M) & local context               & 56.59 & 80.44     & 512                 \\
           &              & +implementation\&usages & 69.28 & 88.08     & 1024                 \\
           & PLBART-base (139M) & local context               & 46.68 & 73.96    & 512                 \\
           &              & +implementation\&usages & 57.25 & 80.94     & 1024                 \\        
           & UnixCoder-base (126M) & local context               & 54.31 & 78.53  & 512                 \\
           &              & +implementation\&usages & 66.20 & 86.05   & 924                 \\           
\bottomrule
\end{tabular}
}
\caption{Performance of different models with task-specific fine-tuning on \CallArgs.
Unidirectional results are grouped by model types: decoder-only (top) and encoder-decoder (middle).
In-filling results are from encoder-decoder models (bottom).
}
\label{tab:fc_all}
\end{table*}

\subsection{RQ3}
\label{sec:ablation-study}
\ifaaai
We conduct an ablation study on the impact of function implementation context and function usage context. We use CodeT5 to study how different types of contexts, and more closely, how the use of usage contexts, would influence the argument completion performance. 
We choose CodeT5 because its pretraining tasks align well with our task, and as evidenced by our results in the previous subsection,  %
CodeT5 achieves the best results across similar model sizes.
The total input length is 512 unless otherwise specified.

\subsubsection{Effect of implementation and usage information.}

\begin{table}[ht]
\centering
\ifaaai
\resizebox{\columnwidth}{!}{
\fi
\begin{tabular}{ccccc}
\toprule
Task       & Context                     & EM   & EditSim & SPM* \\
\midrule
Unidirectional & local context           & 47.16 & 74.44 & 89.22 \\
               & +implementation         & 52.66 & 78.66 & 97.91 \\
               & +usages                 & 58.95 & 81.99 & 94.76 \\
               & +implementation\&usages & 61.59 & 83.73 & 98.64 \\
\midrule
In-filling     & local context           & 56.59 & 80.44 & 91.8  \\
               & +implementation         & 60.22 & 82.97 & 98.27 \\
               & +usages                 & 65.57 & 85.86 & 95.88 \\
               & +implementation\&usages & 67.59 & 87.10 & 98.77 \\
\bottomrule
\end{tabular}
\ifaaai
}
\caption{Performance of CodeT5-base with different auxiliary contexts. }
\label{tab:fc_result}
\fi
\end{table}

\Tabref{tab:fc_result} shows the completion results with different auxiliary contexts. We find that both function implementation and function usage information are beneficial for call argument completion. Adding usage information leads to higher performance gain. 

We also report in \Tabref{tab:fc_result} Surface-level Positional Matching (SPM*), which checks the rate where the predicted arguments can match the parameters in the function definition.
We see that function implementation is more helpful in improving SPM*, suggesting the importance of accessing function definition in getting the number of arguments and the keyword prefixes right.

\subsubsection{Effect of the number and length of function usage contexts.}

\begin{table}[ht]
\centering
\small
\begin{tabular}{ccccc}
\toprule
Task           & EM   & EditSim & Input length & Usages \\
\midrule
Unidirectional & 61.59 & 83.73 & 512 & (3, 64)  \\
               & 62.73 & 84.33 & 1024 & (3, 128) \\
               & 63.00 & 84.46 & 1024 & (6, 64) \\
               & 62.87 & 84.46 & 1024 & (8, 64) \\
\midrule
In-filling     & 67.59 & 87.10 & 512 & (3, 64)  \\
               & 69.28 & 88.08 & 1024 & (3, 128) \\
               & 69.26 & 88.02 & 1024 & (6, 64) \\
               & 69.00 & 87.92 & 1024 & (8, 64) \\
\bottomrule
\end{tabular}
\caption{Performance of CodeT5-base when using different numbers and lengths of usage contexts. ``Usages'' column indicates the number of function usages used and the average length budget for each usage.}
\label{tab:fc_length}
\end{table}

\begin{table}[ht]
\centering
\small
\begin{tabular}{cccc}
\toprule
Task       & Context                     & EM   & EditSim \\
\midrule
Unidirectional & +implementation         & 52.66 & 78.66 \\
               & w/ usages (threshold)   & 56.26 & 80.54 \\
               & w/ usages (CDI)         & 57.61 & 81.12 \\
               \cmidrule{2-4}
               & +implementation\&usages & 61.59 & 83.73 \\
\midrule
In-filling     & +implementation         & 60.22 & 82.97 \\
               & w/ usages (threshold)   & 62.55 & 84.40 \\
               & w/ usages (CDI)         & 64.24 & 85.16 \\
               \cmidrule{2-4}
               & +implementation\&usages & 67.59 & 87.10  \\
\bottomrule
\end{tabular}
\caption{Comparing using function usage information during training and during inference for CodeT5-base. }
\label{tab:fc_copy}
\end{table}

We vary the number and the length budget of function usages used. 
The results are shown in Table \ref{tab:fc_length}. 
We find that longer context helps as enlarging the model input length from 512 to 1024 improves the argument completion performance.
Using more function usage information does not bring significant improvements when the total input length is fixed. 
This may be because the coverage of exact or similar usages would saturate as the number of usages increases as we can see in Appendix \ref{sec:saturate}. 
A better way to leverage more usage information is a meaningful future direction.

\subsubsection{Usage copying.}
One concern is if the models are simply copying the arguments from other usages. 
Therefore, we evaluate the performance of copying the top usage if the similarity is above a certain threshold. Otherwise, the model output is used. We set the threshold using the validation set. 
In this experiment, the model is fine-tuned with implementation context but not usage context.
We also check whether concatenating the usage information at inference directly (CDI) would give similar performances. 
The result is shown in Table \ref{tab:fc_copy}. 

We find that the threshold copying indeed improves the prediction, which confirms the existence and merit of exact matches. 
However, the model can better leverage additional patterns and relations (CDI) than simple copying (threshold).
On the other hand, using function usage information during the model training (+implementation\&usages) brings the most performance gain. 
It indicates the nontrivial ability of our best performing models to attend usage examples and compose appropriate arguments from them.

\fi
\ifamlc

Due to space limit, we summarize findings here and leave detailed results and discussions in \Appref{app:ablation-study}.
We find that
\begin{enumerate*}[label=\itshape\roman*\upshape)]
    \item both function implementation and function usage information are beneficial for call argument completion, where function definition is particularly beneficial in getting the number of arguments and the keyword prefixes right;
    \item increasing input length from 512 to 1024 improves the argument completion, while increasing the number of function usages alone does not bring significant improvements;
    \item similarity-based copying also improves call argument completion against local-context baselines, which confirms the existence and merit of exact matches, although the performance gap is clear against the best performing models.
\end{enumerate*}

\fi

\section{Related Work}
\label{sec:related-work}

\paragraph{Datasets.}
The lack of cross-file and cross-project (e.g. dependencies) information is a general issue in current evaluation datasets for code.
In terms of code completion, common choices are Py150 \citep{raychev2016probabilistic} for Python and Github Java Corpus \citep{allamanis2013mining} for Java. Both datasets are constructed at file level, where source files are isolated from their project and dependencies and no consideration of project separation is taken in constructing training and test sets.
\citet{lu2022reacc} constructed a code completion dataset from CodeNet \citep{puri2021project}, which contains coding problems and solutions from online judge websites and also lacks project context. 
\citet{clement2021long} presented a real-world Python method generation task based on CodeSearchNet \citep{husain2019codesearchnet} but the auxiliary information they extract still comes from within a local file. 
\citet{svyatkovskiy2021fast} constructed a completion dataset based on top Python repositories on GitHub and released the URLs for these repositories. 
However, those repositories are not write-protected and can change over time. Besides, setting up the dependency environments at scale for further analysis is not easy. 
Both make their dataset difficult to reproduce.
In the contrast, we release the code and the dependencies for the projects to ensure reproducibility.
Apart from code completion, datasets for other code tasks such as Cloze test \citep[e.g.][]{feng2020codebert}, code refinement \citep[e.g.][]{tufano2019empirical, yasunaga2021break, haque2022fixeval}, and generating code from text descriptions \citep[e.g.][]{chen2021evaluating, hendrycks2021measuring, austin2021program}, are often small and mostly without project-level code context. 
Beyond-local information is beneficial for programmers to solve programming tasks in real-world settings. The lack of such information in the current dataset would restrict the progress into high-level semantic understanding and reasoning in the code domain.

\paragraph{Code language models.}
Encouraged by the success of pretrained language models in natural language processing \citep{devlin2019bert, liu2019roberta, lewis2019bart, raffel2020exploring} and the promise of naturalness in code \citep{hindle2016naturalness, allamanis2018survey}, we have seen rising adaptations of language models for code. For example, CuBERT \citep{kanade2020learning} and CodeBERT \citep{feng2020codebert} are pretrained based on masked language modeling. GPT-C \citep{svyatkovskiy2020intellicode} and CodeGPT \citep{lu2021codexglue} are both pretrained based on unidirectional language modeling. PLBART \citep{ahmad2021unified} and CodeT5 \citep{wang2021codet5} are pretrained encoder-decoder structures which adopts denoising objectives and can support code understanding and code generation. UnixCoder \citep{guo2022unixcoder} combines the above three pretraining objectives for a unified pretrained model.

\paragraph{Code completion.}
Code completion is an essential feature for modern IDEs and an important topic for code intelligence. 
In recent years, deep neural networks \citep{liu2016neural, li2018code, alon2020structural, liu2020multi, kim2021code}, especially pretrained language models \citep{svyatkovskiy2020intellicode, lu2021codexglue} become the mainstream solution to this task. 
Still, incorporating additional information proved beneficial.
One popular choice is abstract syntax tree, e.g. \citet{kim2021code, peng2021could, guo2022unixcoder}. 
However, \citet{lopez2022ast} suggested that pretrained code language models may have already encoded the syntax.  
Other proposals seek to use data flow graph, control graph, and various graph relations, e.g. \citet{guo2020graphcodebert, hellendoorn2019global}.
However, information is still restricted from a single file.
We instead try to enhance the model with out-of-file information, similar to what is accessible in a development environment.

For project-level analyzer induced information, \citet{svyatkovskiy2021fast} described a way to use a static analyzer to refine completion candidates from neural methods.
\citet{weyssow2020combining} considered leveraging the project-wise contexts via embeddings for better function call completion performance.
Other than code completion, project-level information has been utilized for methods name prediction~\citep{wang2021lightweight} and generating code from text descriptions~\citep{lyu2021embedding}.
However, none of them tested their approaches with pretrained code language models. 
In terms of incorporating additional context through concatenation,
\citet{clement2021long} reported improvements from prioritize certain parts of in-file context.
Recently, \citet{lu2022reacc} proposed to enhance code language models by concatenating similar code fragments retrieved by a neural network. Despite the general similarity, we 1) use a simple lightweight way to retrieve auxiliary information instead of training a heavy retriever; 2) do not restrict ourselves on similar code fragments and show that dissimilar code fragments (function implementation) can be helpful; 3) explore task-specific fine-tuning with retrieved information for better completion.

\section{Conclusion}
We curated \PyEnvs, a collection of permissively licensed Python packages along with isolated development environments. 
Upon that, we built a function call argument completion dataset \CallArgs containing analyzer-induced information for each function call. 
We experimented feeding auxiliary information as additional input context to various pretrained code language models for call argument completion during training and inference.
Results show that access to the function implementation and function usages universally improves the model performances.
We further provide insights on the effect of different types of models and different types of additional information on this task.
In the future, we can use \PyEnvs to construct new datasets for other code-related tasks to further study the benefits from cross-file and cross-project information.

\bibliography{main}

\clearpage

\appendix
\section{Details of Dataset Curation}
\label{app:dataset_details}

\subsection{Requests to language server}
\label{sec:query}

\begin{table*}[ht]
\centering
\begin{tabular}{ll}
\toprule
File content  \\
\midrule
\begin{tabular}[c]{l}
\begin{lstlisting}
import torch
a = torch.zeros(5)
\end{lstlisting}
\end{tabular} \\
\midrule
Query & Response \\
\midrule
\begin{tabular}[c]{l}
\begin{lstlisting}
{
  "jsonrpc": "2.0",
  "id" : ...,
  "method": "textDocument/definition",
  "params": {
    "textDocument": {
      "uri": "file://path/to/file/"
    },
    "position": {
      "line": 1, # start from 0
      "character": 10
    }
  }
}
\end{lstlisting}
\end{tabular} &
\begin{tabular}[c]{l}
\begin{lstlisting}
{
  "jsonrpc": "2.0",
  "id" : ...,
  "result": [{
    "uri": "file://path_to_lib/torch/ \
      _C/_VariableFunctions.pyi",
    "range":{
      "start": {
        "line":1547,
        "character":4
      },
      "end": {
        "line":1547,
        "character":9
      }
    }
  }]
}
\end{lstlisting}
\end{tabular}
\\
\bottomrule
\end{tabular}
\caption{Example request and response for resolving the definition location of \codeinline{zeros}. The header part is omitted.}
\label{tab:ls_examples}
\end{table*}

Usually, development tools and IDE communicate with the server using the language protocol over JSON-RPC. 
The base protocol consists of a header and a content part. The header part consists of two fields: ``Content-Length'' which denotes the length of the content part in bytes, and ``Content-Type'' which denotes the MIME type of the content part. 
The content part describes requests, responses and notifications for the language server. 
The content part of a request usually contains ``method'' and ``params''  fields which denote the method of the language server to be invoked and its parameters. 
For each request, the language server must return a response, containing the result of the request or an error message. 
Table \ref{tab:ls_examples} shows an example of the request and response for resolving the definition location of a symbol (textDocument/definition). 
More details about the formats of other language capabilities can be found in the official website \footnote{https://microsoft.github.io/language-server-protocol/specifications/lsp/3.17/specification/}.

\subsection{Selection of target function calls}
\label{sec:criteria}

We use the following criteria to select function calls for constructing the function call argument completion dataset.
\begin{enumerate}
    \item Ignore function calls related to error messages and logging, e.g. \codeinline{raise VelueError(...)} and \codeinline{print(...)}.
    \item Ignore function calls related to standard Python type constructions or conversions, e.g. \codeinline{int} and \codeinline{dict}.
    \item Ignore function calls related to unit test, which usually starts with \codeinline{assert}.
    \item Ignore function calls that occur outside of any function body.
    \item Ignore function calls with no argument. They may make the task too easy.
    \item Ignore function calls whose definitions cannot be found by the language server.
    \item Ignore function calls whose ground truth arguments contains string literals. 
    \item Ignore other special functions like \codeinline{sleep} and \codeinline{add\_argument}.
\end{enumerate}

\subsection{Information isolation} 
\label{sec:isolation}

One drawback of many existing dataset for code completion is that they do not provide or are not careful about project-level isolate between their training and test split. See discussion in \Secref{sec:related-work}.
For a set of test projects $t$ and a set of training projects $s$, we define the  sets containing direct imports $t_1 = t \cup \{p_1 \mid \exists p \in t, p~\text{imports}~p_1\}$ and $s_1 = s \cup \{p_1 \mid \exists p \in s, p~\text{imports}~p_1\}$.
We consider four levels of project isolation:
\begin{enumerate}[label={Level~\arabic*:},leftmargin=0em,itemindent=4em]
    \item $s \cap t = \emptyset$. This is the project-level isolation.
    \item $s_1 \cap t = \emptyset$. It guarantees that the training projects would not see the information about the test projects.
    \item $s_1 \cap t = \emptyset, s \cap t_1 = \emptyset$. It further guarantees that the test projects would not leverage the information about the training projects.
    \item $s_1 \cap t_1 = \emptyset$. No information is shared between the training projects and test projects through their direct imports.
\end{enumerate}
Because function implementation often come from dependent packages, we enforce the strictest level-4 isolation when making the train/valid/test split for \CallArgs. 
We first randomly sample a set of test projects and then select training projects that satisfy the isolation constraint.
Read more about \CallArgs in  \Secref{sec:dataset-callargs}.

We deem this level of isolation sufficient as we extract no information from beyond direct dependencies for the task of function call argument prediction.
Enforcing isolation for all transitive dependencies would further reduce the number of available projects for \CallArgs, which is currently 1868 out of the 2814 available projects. 
We will release the full set of \PyEnvs packages along with the scripts used to create \CallArgs. 
Future work can thus apply different isolation strategies if it relies on incorporating information from transitive dependencies.

\subsection{Context example}
\begin{table*}[ht]
\centering
\begin{tabular}{p{\textwidth}}
\toprule
Local context. \codeinline{<PREDICT>} denotes the prediction location.  \\
\midrule
\begin{tabular}[c]{l}
\begin{lstlisting}[]
def _set_arguments(self,arguments,context):
    positional, named = arguments
    variables = context.variables
    args, kwargs = self.arguments.map(
\end{lstlisting}
\end{tabular} \\
\ContinueLineNumber
\begin{tabular}[c]{l}
\begin{lstlisting}
        <PREDICT>
\end{lstlisting}
\end{tabular} \\
\ContinueLineNumber
\begin{tabular}[c]{l}
\begin{lstlisting}[]
    self._set_variables(args, kwargs, variables)
    context.output.trace(lambda: self._trace_log_args_message(variables))
\end{lstlisting}
\end{tabular} \\
\midrule
Function implementation context. \\
\midrule
\begin{tabular}[c]{l}
\begin{lstlisting}
def map(self, positional, named, replace_defaults=True):
    mapper = ArgumentMapper(self)
    return mapper.map(positional, named, replace_defaults)
\end{lstlisting}
\end{tabular} \\
\midrule
One function usage context. \\
\midrule
\begin{tabular}[c]{l}
\begin{lstlisting}
def resolve_arguments(self, arguments, variables=None):
    positional, named = super().resolve_arguments(arguments, variables)
    if not self._supports_kwargs:
        positional, named = self.arguments.map(positional, named)
    return positional, named
\end{lstlisting}
\end{tabular} \\
\bottomrule

\end{tabular}
\caption{An argument completion example for \codeinline{self.arguments.map()}. }
\label{tab:full}
\end{table*}

In Table \ref{tab:full}, we give an example of the function implementation context and a function usage context extracted for the instance shown in Table \ref{tab:examples}.

\section{Details of Experiments}

\subsection{Implementation environment} 
We implement our code based on Pytorch \footnote{https://pytorch.org/} and Transformers \footnote{https://github.com/huggingface/transformers}. The Python version for implementing the codes and constructing the dataset is Python 3.9.12. All pretrained code language models are automatically downloaded from Hugging Face \footnote{https://huggingface.co/models}. 

\subsection{Performance on the different types of the function calls}
\begin{table*}[ht]
\centering
\begin{tabular}{ccccccccc}
\toprule
\multirow{2}{*}{Task} & \multirow{2}{*}{Model} & \multirow{2}{*}{Context}  & \multicolumn{2}{c}{stdlib} & \multicolumn{2}{c}{in-project} & \multicolumn{2}{c}{third-party} \\
\cmidrule{4-9}
                      &                        &                          & EM      & EditSim   & EM    & EditSim        & EM         & EditSim \\
\midrule
Unidirectional        & CodeGen-MONO          & local context.            & 51.81   & 77.58     & 46.32 & 73.67           & 34.23         & 67.62   \\
                      &                        & +implementation\&usages  & 60.00   & 81.49     & 66.02 & 86.42           & 52.52         & 77.00   \\
                      & CodeT5-base            & local context            & 52.38   & 77.57     & 45.02 & 73.21           & 35.77         & 67.30   \\
                      &                        & +implementation\&usages  & 60.12   & 81.99     & 66.53 & 87.20           & 50.86         & 77.20   \\
\midrule
In-filling            & CodeT5-base            & local context            & 61.48   & 82.97     & 54.76 & 79.60           & 44.77         & 73.73   \\
                      &                        & +implementation\&usages  & 68.27   & 86.78     & 71.89 & 90.06           & 57.39         & 81.61   \\
\bottomrule
\end{tabular}
\caption{The performance of different types of the function calls on our new call argument completion dataset.}
\label{tab:types}
\end{table*}

The function definitions of the calls in our dataset may come from three sources: 
1) other files in the same project (in-project); 
2) third-party dependencies (third-party); and
3) the built-in standard libraries of Python (stdlib). 
We evaluate the model performance on these three types of functions using the same models from Table \ref{tab:fc_all}. 
The results are shown in Table \ref{tab:types}. 

We find that when using the local context only, the models perform best on the function calls defined in the standard libraries. 
This may be partly due to that the standard libraries are treated as the shared information between the training and test set, and partly due to that this type of functions constitutes a relative large amount of occurrences. 
Both of the facts make it easier for the models to pick up and generalize the patterns even if only the local context is available.

However, when using function implementation context and function usage context, the completion performance for in-project and third-party function calls greatly increased. 
Especially for in-project function calls, now their EM and EditSim numbers even surpass those of stdlib calls.
It provides yet another evidence for the importance of using in-project and cross-project information in models' generalizability cross projects for code tasks. 

\subsection{Exact match in the usages} 
\label{sec:saturate}
\begin{figure*}[bt]
\centering 
\includegraphics[width=0.5\linewidth]{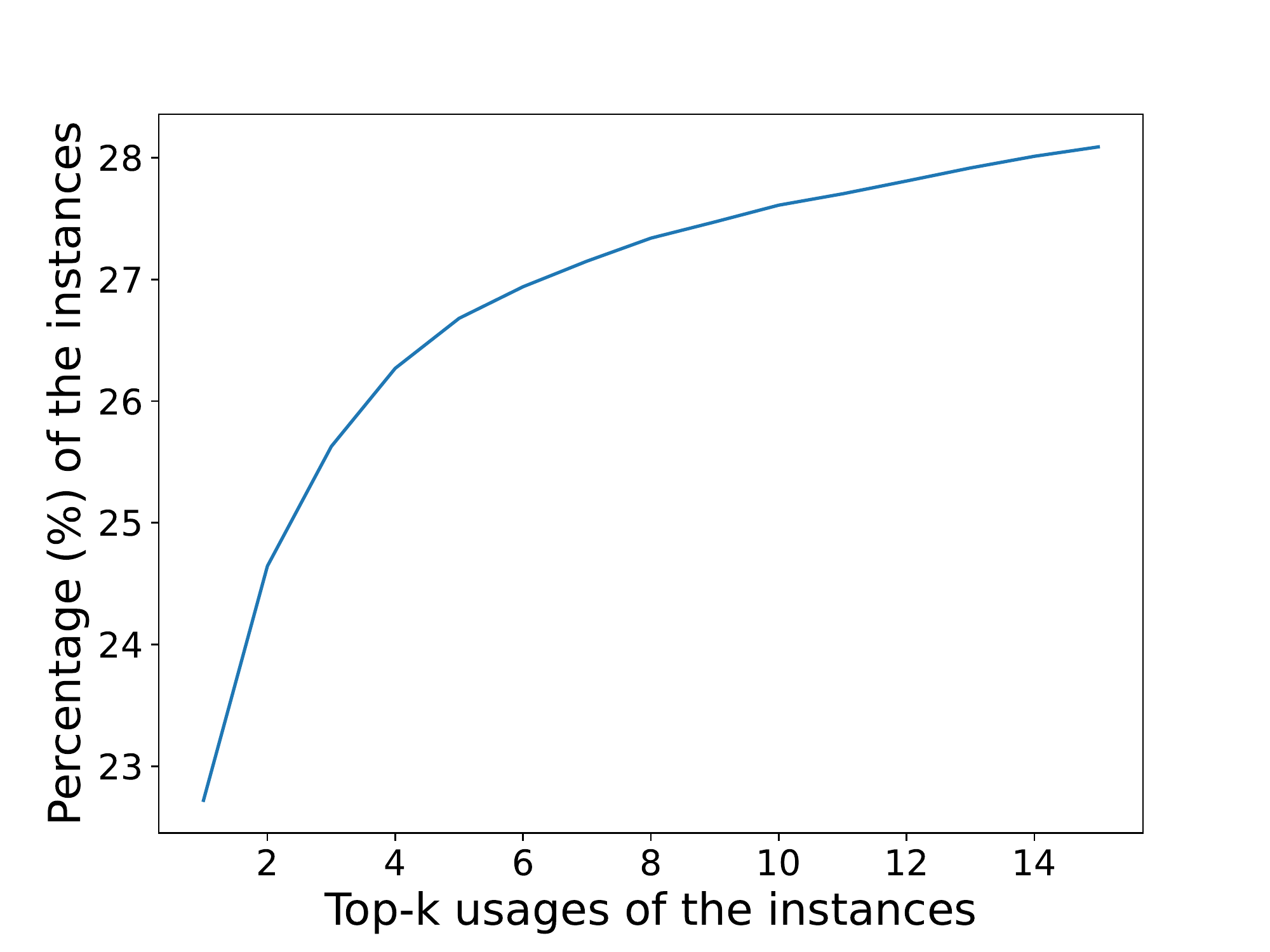}
\caption{Percentage of function calls where at least one of their top-k usages shares the same arguments. Calculated from the \CallArgs validation set.}
\label{fig:curve}
\end{figure*}

We see from \Figref{fig:curve} that the coverage of exact argument matches in top-$k$ usages increases not much when $k > 6$. 
It partially explains the observations in \Tabref{tab:fc_length} that providing more numbers of usages does not necessarily bring benefit to the model performance.

\end{document}